\documentclass[preprint,pre]{revtex4}
\usepackage{epsfig}
\usepackage{floatflt}
\usepackage{graphicx}
\usepackage{amsmath}
\usepackage{setspace}
\begin{document}

\title{Study of the heating effect contribution to the nonlinear dielectric response of a supercooled liquid}

\author{C. Brun$^1$}
\author{C. Crauste-Thibierge$^2$}
\author{F. Ladieu$^{1 \star}$}
\author{D. L'H\^ote$^{1 \star}$}
\email{francois.ladieu@cea.fr, denis.lhote@cea.fr}

\affiliation{$^1$ SPEC (CNRS URA 2464), DSM/IRAMIS CEA Saclay, Bat.772, F-91191 Gif-sur-Yvette  France}
\affiliation{$^2$ LLB (CNRS UMR 12), DSM/IRAMIS CEA Saclay, Bat. 563, F-91191 Gif-sur-Yvette  France}

\date{\today}

\begin{abstract}
We present a detailed study of the heating effects in dielectric measurements carried out on a liquid. 
Such effects come from the dissipation of the electric power in the liquid and give a contribution to the nonlinear 
third harmonics susceptibility $\chi_3$ which depends on the frequency and temperature. 
This study is used to evaluate a possible `spurious' contribution to the recently measured nonlinear susceptibility of an 
archetypical glassforming liquid (Glycerol) . Those measurements have been shown to give a direct evaluation of the 
number of dynamically correlated molecules temperature dependence close to the glass transition temperature 
$T_g \approx 190$K (Crauste-Thibierge \textit{ et al.}, Phys. Rev. Lett 104,165703(2010)). We show that the 
heating contribution is totally negligible (i) below $204$K at any frequency; (ii) for any temperature at the 
frequency where the third harmonics response $\chi_3$ is maximum. Besides, this heating contribution does \textit{not} 
scale as a function of $f/f_{\alpha}$, with $f_{\alpha}(T)$ the relaxation frequency of the liquid. In the high 
frequency range, when $f/f_{\alpha} \ge 1$, we find that the heating contribution is damped because the dipoles 
cannot follow instantaneously the temperature modulation due to the heating phenomenon. An estimate of the magnitude 
of this damping is given.
\end{abstract}

\maketitle

\section{\label{I}Introduction}

Upon fast enough cooling, most liquids avoid cristallization and enter a supercooled state. The latter is characterized by the ``viscous slowing down'' phenomenon \cite{Debenedetti01}, an extremely fast increase of the viscosity when the temperature $T$ is decreased. Below the glass transition temperature $T_g$ the viscosity is so high that the system is, in practice, a solid. The glass transition is a longstanding issue of condensed matter physics, since no structural signature has ever been detected around $T_g$, e.g. the static neutron spectra do not change at $T_g$, contrary to what happens for the cristallization \cite{Debenedetti01}. Significant progresses in understanding the physics of the glass transition were made in the last fifteen years, when the heterogeneous nature of the dynamics of supercooled liquids was established, through various experimental \cite{Schiener96, Tracht98, Ediger00, Richert02} as well as numerical works \cite{Hurley95}. 
The important concept of `dynamical heterogeneities' has emerged \cite{Ediger00, Richert02}, according to which the relaxation comes from the collective motion of groups of $N_{corr}$ molecules. All these groups evolve in time, some of them being faster or slower than the average dynamics. The viscous slowing down would come from the fact that this number $N_{corr}$ increases as $T$ decreases towards $T_g$. The temperature dependence of $N_{corr}$ thus became a crucial issue in the field, triggering new theoretical ideas \cite{Berthier05, Berthier07, Dalle07, Bouchaud05}. One of them came from an analogy \cite{Bouchaud05} with the well known spin glass physics in which the nonlinear suceptibility diverges at the critical temperature $T_c$, reflecting the long range amorphous order which sets in at $T_c$. It was proposed in Ref. \cite{Bouchaud05}, that a similar effect occurs in supercooled liquids, with the key difference that the peak of the nonlinear response $\chi_3(\omega)$ should appear at finite frequencies $\omega$ of the order of $1/\tau(T)$ where $\tau(T)$ is the relaxation time of the supercooled liquid. For the first time, a (nonlinear) susceptibility was directly related to the quantity $N_{corr}$ of interest, allowing to scrutinize its temperature dependence. We have performed recently the corresponding experiment for the nonlinear dielectric susceptibility of glycerol \cite{Thibierge08,Crauste10} and shown that the $N_{corr}(T)$ dependence was indeed an increase as $T$ decreases.
              
The nonlinear susceptibility $\chi_3$ measured in Refs. \cite{Thibierge08,Crauste10} corresponds to the detection of the third harmonics of the polarisation $P_{3}$, at three times the frequency of the applied electric field 
$E \cos(\omega t)$. For $E\sim 1$ MV/m, one finds that $P_{3}/P_{1}$ is typically around $10^{-5}$, with $P_{1}$ the polarisation at the frequency $\omega$ (mainly dominated by the linear response of the system). Due to the small value of $P_{3}/P_{1}$, a thorough analysis is required to see whether some `spurious' effects
can affect the $P_{3}$ measurements. Expanding on the arguments given in Ref. \cite{Epaps10}, we study in 
this paper the contribution of the ac heating of the liquid to our $P_{3}$ measurements. 
This heating comes from the fact that the strong applied electric field leads to some dissipated electrical power $p$ with a d.c. component as well as a component $p_{2}(t)$ oscillating at $2 \omega$. Before being absorbed by the thermal reservoir which sets the base temperature $T$ of the experiment (hereafter, the 
metallic electrodes), the heat has to travel across the sample of thickness $e$. Thus the 
dissipated power leads to a small temperature increase, containing a $2 \omega$ component $\delta T_{2} \sim 
p_{2}$. From the first order estimate $\delta P(t) \simeq (\partial P_{1}(t)/ 
\partial T) \delta T(t)$ which involves a product of two terms oscillating at $\omega$ and $2\omega$, one sees that a spurious contribution to the measured $P_{3}$ comes from $\delta T_{2}(t)$, i.e. from a part of the heating phenomenon.

We shall see below that, close to $T_g$, this heating contribution to $P_{3}$ vanishes in the limit of extremely thin samples ($e \to 0$). 
This is an illustration of the `spurious' nature of this heating contribution. We emphasize that the heating we study here must not be confused with the `heterogeneous heating' assumed in a model put forward recently by Richert 
\textit{et al.} \cite{Richert06,Richert07a,Richert08}. This `heterogeneous heating model' gives a 
phenomenological description of the intrinsic nonlinear effects in supercooled liquids, and was shown to account 
for nonlinear experimental data at $\omega$ in Ref. \cite{Richert06}. In this model, each 
dynamical heterogeneity has its own fictitious temperature `on top' of the temperature of the phonon bath. 
At the present time, it is the only model which allows to calculate the nonlinear susceptibility near $T_g$ and we intend to give a thorough study of its predictions at $3 \omega$ in a future paper. 
To emphasize the difference between this heterogeneous heating model and the study of the present paper, 
we call the heating contribution to the nonlinear susceptibility investigated here 'homogeneous heating contribution'.  

The paper is organised as follows: in section \ref{II} we derive the relations which allow to calculate the heating contribution to $P_{3}$. In section \ref{III} we give the corresponding results, and clearly show the temperature and frequency ranges where the heating contribution is negligible in our experiments. We eventually show that in the high frequency range, i.e. when $\omega \tau(T) \gg 1$, the heating contribution is damped by the finite relaxation time $\tau(T)$ of the dipoles, and we estimate the magnitude of this damping.

\section{\label{II} Heating effects calculations}

In section \ref{IIA} we establish the link between the temperature increase and the homogeneous heating contribution to $P_3$. The detailed calculation of the temperature increase is postponed to section \ref{IIB}.

\subsection{\label{IIA} From $\delta T$ to the homogeneous heating contribution to $\chi_3$.}

This subsection is divided into three parts: we first establish an upper bound of the heating contribution to $P_3$. We then move to an estimate of the damping of this contribution at high frequencies, due to the finite relaxation time of the dipoles. Last, we show what is the most natural quantity to plot to compare the heating contribution to $P_3$ to the results of Ref. \cite{Crauste10}.

\subsubsection{\label{IIA1} The heating contribution without damping}

Let us consider a sample made of a supercooled liquid excited by an oscillating field $E \cos(\omega t)$. For small enough {\it E} values, the resulting linear polarisation $P_{lin}$ reads :
\begin{equation}\label{E1}
\frac{P_{lin}(t)}{\epsilon_0 E}= \chi_{1}' \cos(\omega t) +\chi_{1}'' \sin(\omega t),
\end{equation}
where $\epsilon_0$ is the dielectric permitivity of vacuum. In general, $\chi_1'$ and $\chi_1''$ strongly depend on frequency $f$ = $\omega/2\pi$. We define the frequency $f_{\alpha}$ which characterizes the relaxation at a given temperature $T$, as the frequency where $\chi_1''$ is maximum. This relaxation frequency 
$f_{\alpha}$ is strongly $T$ dependent, and is of the order of $1/\tau$ where $\tau(T)$ is the average relaxation time of the dipoles of the supercooled liquid. 

As evoked above, the volumic density of dissipated electrical power $p(t)$ contains a d.c. term and a term oscillating at $2\omega$ (see below Eq.~(\ref{E14})). The resulting heat propagates towards the `thermostat'. The resulting average sample temperature increase $\delta T(t)$ can be written: 

\begin{equation}\label{E2}
\delta T(t) = \delta T_{0} +\delta T_{2} \cos(2\omega t - \phi_2),
\end{equation}
where the mean dc temperature increase $\delta T_0$ is larger than or equal to the ac one $\delta T_2$, thus at any time $\delta T(t) \ge 0$. $\phi_2$ is a phase shift related to heat transport that will be given in section \ref{IIB}. As our measurements \cite{Crauste10} give the nonlinear dielectric response averaged over the sample volume, $\delta T(t)$ in Eq.~(\ref{E2}) is the temperature increase averaged over the same volume. Using Eqs.~(\ref{E1}),(\ref{E2}), we thus obtain for the nonlinear part of the polarisation due to heating effects:

\begin{equation}\label{E3}
\frac{P(t) - P_{lin}(t)}{\epsilon_0 E}= \left( \frac{\partial \chi_{1}'}{\partial T} \delta T(t) \right) \cos(\omega t) +\left( \frac{\partial \chi_{1}''}{\partial T} \delta T(t) \right) \sin(\omega t).
\end{equation}

This is an upper limit of the heating contribution to the nonlinear response, since we have assumed that $\delta T(t)$ induces instantaneously a modification of the susceptibility. As already advocated in Ref. \cite{Richert08}, this is questionable, specially in what concerns the contribution of $\delta T_2(t)$ which should be damped because of the finite relaxation time $\tau$ of the dipoles. This point is adressed in the next section \ref{IIA2}. From Eqs.~(\ref{E2}) and (\ref{E3}), one gets :

\begin{eqnarray}
 \frac{P(t) - P_{lin}(t)}{\epsilon_0 E} & = & \left[\left(\delta T_0 + \frac{1}{2}\delta T_2 \cos(\phi_2)\right)\frac{\partial \chi_{1}'}{\partial T} + \frac{1}{2}\delta T_2 \sin(\phi_2)\frac{\partial \chi_{1}''}{\partial T}\right] \cos(\omega t)\nonumber \\
   &   & + \left[\left(\delta T_0 - \frac{1}{2}\delta T_2 \cos(\phi_2)\right)\frac{\partial \chi_{1}''}{\partial T}  + \frac{1}{2}\delta T_2 \sin(\phi_2)\frac{\partial \chi_{1}'}{\partial T}\right] \sin(\omega t)\nonumber \\
&   & + \left[ \frac{1}{2}\delta T_2 \cos(\phi_2)\frac{\partial \chi_{1}'}{\partial T} - \frac{1}{2}\delta T_2 \sin(\phi_2)\frac{\partial \chi_{1}''}{\partial T}\right] \cos(3\omega t)\nonumber \\
&   & + \left[ \frac{1}{2}\delta T_2 \sin(\phi_2)\frac{\partial \chi_{1}'}{\partial T} + \frac{1}{2}\delta T_2 \cos(\phi_2)\frac{\partial \chi_{1}''}{\partial T}\right] \sin(3\omega t).
\label{E4}
\end{eqnarray}

The four terms in the right hand side of Eq.~(\ref{E4}) give the nonlinear response of the system due to homogeneous heating: 
it contains two terms oscillating at $\omega$ that we shall disregard since they contribute to the nonlinear part of $P_{1}$. 
We shall only keep the two terms of Eq.~(\ref{E4}) oscillating at $3\omega$ to obtain the heating contribution $P_{3,h}$ to 
the third harmonics $P_3$. As we are interested in the heating contribution $\chi_{3,h}$ to the nonlinear \textit{susceptibility} 
we define, as in Eqs.~(4)-(5) of Ref. \cite{Thibierge08} : 

\begin{equation}\label{E5}
 \frac{P_{3,h}(t)}{\epsilon_0 E} =  \frac{E^2}{4} \chi_{3,h}' \cos(3 \omega t)+ \frac{E^2}{4} \chi_{3,h}'' \sin(3 \omega t).
\end{equation}

As a result, $\chi_{3,h}'$  and $\chi_{3,h}''$ should not depend on $E$: their expression is given by identification with the two last terms of Eq.~(\ref{E4}) and defines, throughout this work, what we call the \textit{overestimated} heating nonlinear susceptibility because it is obtained by neglecting the damping evoked above.

\subsubsection{\label{IIA2} Damping of heating contribution: an estimate}

We now move to the problem evoked above, namely the fact that the finite relaxation time $\tau$ of the dipoles 
which contribute to the dielectric susceptibility should damp the modification of this susceptibility due to the oscillating $\delta T_{2}(t)$, specially in the case $\omega \tau \ge 1$. As a consequence, the heating contribution to $\chi_{3,h}$ should be multiplied by a complex factor $R(\omega \tau)$, with a modulus $\left|R(\omega \tau)\right|$ which is expected to be lower than 1. For a precise calculation of $R(\omega \tau)$, one should replace Eq. (\ref{E3}) by an equation accounting for the dynamics of the dipoles in the case of a thermal bath where the temperature has an oscillating component, which is of great complexity. For an estimate, we make two very simplifying assumptions: 

(i)  We assume that the dipoles have a Debye dynamics with a given characteristic time $\tau(T)$. This is a simplifying assumption in the sense that when $T$ is close to $T_g$, it is well known that $\chi_1(\omega)$ is ``stretched'' with respect to a simple Debye law. In fact the Debye dynamics holds only at much higher temperatures, where the molecular motions are independent of each other, which allows to describe the non inertial rotational Brownian motion by the Smoluchowski equation for the probability distribution function of the orientations of the dipoles in configuration space \cite{Dejardin00,Langevin}. After an ensemble averaging of this equation, one gets the well known Debye equation for the dynamics of the average polarisation $P$ \cite{Dejardin00} :
\begin{equation} \label{E8}
\tau \frac{\partial P}{\partial t} + P = \epsilon_{0} \Delta \chi_1 E \cos(\omega t) , 
\end{equation}

where $\Delta \chi_1$ = $\chi_1(\omega = 0) - \chi_1(\omega \rightarrow \infty)$ is the part of the static linear 
susceptibility corresponding to the slow relaxation process we consider.

(ii) We assume that the main effect of the temperature variation $\delta T_2(t)$ is to modulate in time the value of $\tau$ while leaving unchanged the (Debye) dynamics. This can be justified by the fact that the temperature oscillation modulates the viscosity $\eta$, thus also the relaxation time $\tau$ which is proportional to $\eta$ \cite{justification}. Considering the temperature variations of Eq. (\ref{E2}), $\tau(t)$ is now given by 

\begin{equation} \label{E9}
\tau(t) = \tau_{lin} + \left(\frac{\partial \tau_{lin}}{\partial T} \right) \delta T_0 + \left(\frac{\partial \tau_{lin}}{\partial T} \right) \delta T_2 (t) ,
\end{equation}
where $\tau_{lin}$ is the value of $\tau$ at zero field. In the following, $\delta \tau_2$ will denote the amplitude of the $2 \omega$ modulation of $\tau$ due to $\delta T_2(t)$ and corresponding to the last term of Eq. (\ref{E9}). Of course, using Eq. (\ref{E9}) for $\tau(t)$ assumes that $\delta T_2(t)$ instantaneously fully affects $\tau$. A thorough modelization of this problem could lead to a more involved expression where $\delta \tau_2$ would be weaker than in the above expression. As we shall find that the third harmonics is proportionnal to $\delta \tau_2 /\tau_{lin}$, see below Eq. (\ref{E11}), we are led to the conclusion that our new estimate should, again, be slightly overstimated.

We now insert $\tau(t)$ in Eq. (\ref{E8}) and set :
\begin{equation} \label{E10}
P(t) = P_{lin} \cos(\omega t - \Psi_{lin}) + \delta P_1 \cos(\omega t - \Psi_1) + P_3 \cos(3\omega t - \Psi_3) + ...
\end{equation}
where $P_{lin}, \Psi_{lin}, \delta P_1, \Psi_1, P_3, \Psi_3$ are to be determined. As we are only interested in the onset of nonlinear effects, $P_{lin}\propto E$ is much larger than $\delta P_1 \propto E^3$ and than  $P_3 \propto E^3$. This allows to neglect higher order harmonics (denoted by the dots in Eq. (\ref{E10})) and to resolve Eq. (\ref{E8}) by identification of the terms which have the same frequency and the same power of $E$. This yields:

\begin{eqnarray}
 P_3& = & \frac{\epsilon_{0} \Delta \chi_1 E}{2}\frac{\delta \tau_2}{\tau_{lin}} \frac{\omega \tau_{lin}}{\sqrt{1+(\omega \tau_{lin})^2}\sqrt{1+(3\omega \tau_{lin})^2}}  \nonumber \\
\Psi_3 & = & \phi_2 + \arctan{(\omega \tau_{lin})} + \arctan{(3\omega \tau_{lin})}+ \frac{\pi}{2} ,
\label{E11}
\end{eqnarray}

where we remind that $\delta \tau_2$ is the amplitude of the $2 \omega$ modulation of $\tau$ due to $\delta T_2(t)$ in the last term of Eq. (\ref{E9}): thus $\delta \tau_2 \propto E^2$, see Eq. (\ref{E7}) below, which yields the expected $P_3 \propto E^3$. 

We now have to compare with the result obtained if we start from Eq. (\ref{E3}) and use a Debye linear susceptibility. 
A straightforward calculation shows that in that case the third harmonics of the polarisation is given by the solution 
of Eqs.~(\ref{E8}-\ref{E10}) divided by the function $R(\omega \tau)$ introduced above. The expression
for $R(\omega \tau)$ is written using the complex notation :

\begin{equation}\label{E12}
R(\omega \tau) =  \frac{\sqrt{1+(\omega \tau_{lin})^2}}{\sqrt{1+(3 \omega \tau_{lin})^2}} 
\times \exp{\left[ i \times \arctan{\left(\frac{-2\omega \tau_{lin}}{1+3(\omega \tau_{lin})^2} \right)} \right]} .
\end{equation}

As expected $R(\omega \tau \ll 1) =1$ (no damping) and the damping arises at high frequencies since we draw from Eq.~(\ref{E12})  $\vert R(\omega \tau \gg 1) \vert < 1$. More precisely $\vert R(\omega \tau \to \infty) \vert =1/3$, which comes from the fact that $\tau(t)$ enters in Eq. (\ref{E8}) as a factor of $\partial P/ \partial t$: this gives, in Eq.  (\ref{E8}),  a weight $3 \omega \tau$ to $P_3$, contrary to the case where one starts from Eq. (\ref{E3}) where this weight is simply $\omega \tau$. Let us note that when the similar analysis is made for $\delta P_1$, by using the second term ($\propto \delta T_0$) of the right hand side of Eq. (\ref{E9}), no reduction is found at any frequency: The solution found for $\delta P_1(t)$ starting with Eq. (\ref{E3}), is exactly the same, in modulus and phase, as the one found by using Eqs. (\ref{E8}-\ref{E9}). This shows that the reduction of the effect of $\delta T_2(t)$ on the polarisation comes from the fact that, in Eq. (\ref{E8}), $\tau$ and $P$ oscillate together in time. Thus, Eq. (\ref{E12}) can be seen as a first estimate of the fact that the dipoles damp the temperature oscillations, this damping being strong at high frequencies, as physically expected. Of course, one could build a much more thorough model of this effect, but the reduction given by Eq. (\ref{E12}) will be shown to be quite realistic with respect to our experimental data (see below section \ref{IIID}). In practice, to compute the \textit{damped} heating contribution to $\chi_3$, we first compute the overestimated contribution defined in section \ref{IIA1} and then multiply by the complex factor $R(\omega \tau)$ defined above in Eq.~(\ref{E12}).

\subsubsection{\label{IIA3} How to single out the anomalous part of the nonlinear response}

Before moving to the calculation of the temperature increase, let us remind the relation that Bouchaud and Biroli predict \cite{Bouchaud05}, on quite general grounds, between $\chi_3$ and $N_{corr}(T)$ -where $N_{corr}(T)$ denotes the $T$-dependent average number of dynamically correlated molecules-. In Ref. ~\cite{Bouchaud05}, one finds the following scaling form for $\chi_3$ :

\begin{equation}\label{E13}
\chi_3 (\omega,T) \approx \frac{\epsilon_0 (\Delta \chi_1)^2 a^3}{k_BT} N_{corr}(T) \, {\cal H}\left(\omega \tau\right),\qquad
\end{equation}
where $a^3$ the volume occupied by one molecule, and 
${\cal H}$ a certain complex scaling function that reaches its maximum at $\omega \tau \sim 1$ and goes to zero {\it both} for small and large arguments. This 'humped' shape of $\left|{\cal H}\right|$ is due to the glassy correlations:
In the `no correlation case' \cite{Epaps10,Dejardin00}, $N_{corr}(T) \, {\cal H}\left(\omega \tau\right)$
in Eq.~(\ref{E13}) should be replaced by a function which reaches its maximum value at $\omega =0$. Thus $\chi_3(\omega,T)$ can always be considered as the product of a general prefactor $\epsilon_0 (\Delta \chi_1)^2 a^3/k_BT$ times a dimensionless term which summarizes the physics of the system. This is why a natural way to express the various contributions $\chi_{3,i}$ ($i$ indicates the kind of contribution) to $\chi_3$ is to divide them by this prefactor. We thus define a normalized nonlinear susceptibility $X_{3,i}$ = $k_BT$$/$$(\epsilon_0 (\Delta \chi_1)^2 a^3)$$\chi_{3,i}$. We shall consider the normalized heating contribution $X_{3,h}$ by
dividing the heating contributions $\chi_{3,h}$ by the prefactor. Clearly, when no heating contribution or any 
other spurious contribution is present, we expect $X_3(\omega,T)$ = $N_{corr}(T)$${\cal H}(\omega\tau)$.

\subsection{\label{IIB} Calculation of the temperature increase}

In this section, we now calculate the expression of $\delta T(t)$ that one has to introduce in Eq.~(\ref{E4}) in order to obtain the nonlinear response in Eq.~(\ref{E5}). 

The supercooled liquid is characterized by its thermal conductivity $\kappa_{th}$ and its specific heat $c$. As in Ref.~\cite{Birge86}, we consider that $c$ is frequency dependent due to the fact that the slow degrees of freedom cannot contribute to $c$ for frequencies much larger than $f_{\alpha}$. For simplicity we neglect the small imaginary part of $c$ \cite{Birge86, Minakov03}, and we consider also that $\kappa_{th}$ depends neither on the frequency nor on the temperature $T$ \cite{Birge86, Minakov03}. Let us define $(x,y)$ as the plane of our copper electrodes \cite{Crauste10,Epaps10}, with $z=0$ for the lower electrode and $z=e$ for the upper one. Due to their very high thermal conductivity and to their large thickness ($6$ mm), the two electrodes can be considered, to a very good approximation, as a thermostat \cite{electrodes}. The temperature increase $\delta \theta (x,y,z,t)$ of the supercooled liquid at point $(x,y,z)$ and time $t$ thus vanishes for $z=0$ and $z=e$. As the diameter $D$ = 2 cm of the electrodes is typically one thousand times larger than $e \sim 20 - 40\mu$m (see \ref{III}), we may consider that $\delta \theta$ does not depend on $(x,y)$. We obtain $\delta \theta(z,t)$ by solving the heat propagation equation:

\begin{equation}\label{E6}
c \frac{\partial \delta \theta(z,t)}{\partial t} = \kappa_{th} \frac{\partial^2 \delta \theta(z,t)}{\partial z^2} + p(t) ,
\end{equation}

where the dissipated power is given by :
\begin{equation}\label{E14}
p(t)=\frac{1}{2} \epsilon_0 \chi_{1}'' \omega E^2\left(1+\cos(2 \omega t - \phi) \right) \hbox{\ with\ } \phi = -\pi + 2\arctan \left(\frac{\chi_1''}{\chi_1'-\chi_1'(\omega \rightarrow \infty)}\right) .
\end{equation}

This expression of $p(t)$, where the fast, non relevant, degrees of freedom contributing to $\chi_1'(\omega \rightarrow \infty)$ are separated from the slow degrees of freedom corresponding to glassy dynamics, deserves some comments. While the prefactor of 
the right hand side of Eq.~(\ref{E14}) is `textbook' knowledge, the expression of $\phi$ is far less obvious (see the Appendix 
of \cite{Richert08}). This phase comes from the fact that the dissipation arises due to the friction of the dipoles with 
the surrounding molecules, and that this friction force ${\cal F}$ is proportional to $\partial P/\partial t$. The simplest 
example is the case of Debye dynamics where $\cal F$ is proportionnal to the first term of Eq.~(\ref{E8}), see \cite{Frolich}. 
Then, the power 
$p$ corresponding to $\cal F$ must be given by ${\cal F} v$ where $v \sim \partial P/\partial t$ is the `speed' of the dipoles. 
It follows that $p \sim (\partial P/\partial t)^2$ and by using Eq.~(\ref{E1}), with the appropriate prefactor for $p$, 
the expression given in Eq.~(\ref{E14}) comes out. Note that if we consider the limit $\chi_1'' \gg \chi_1'-\chi_1'(\omega \rightarrow \infty)$, we get $\phi \to 0$. This is the case of a metal where the response is not due to dipoles, but to electrons motion, and for
which it is well known that there is no dephasing between 
$p$ and $E^2$. On the other hand, in the limit $\chi_1'' \ll \chi_1'-\chi_1'(\omega \rightarrow \infty)$, which happens in liquids when    
$f \ll f_{\alpha}$, one finds that the $2 \omega$ component of $p$ and of $E^2$ are in phase opposition. In the case of a Debye 
dynamics, Eq.~(\ref{E14}) can be rigourously derived \cite{Richert08} because the expression of the friction force 
is explicit. When the dynamics does not follow the Debye 's law, which is the case of supercooled liquids close to $T_g$, the 
dynamical equation obeyed by each dynamical heterogeneity is not known \cite{Richert02}. However, Eq.~(\ref{E14}) should 
still remain valid since the assumption ${\cal F} \sim \partial P/\partial t$ amounts to the lowest order development of 
the general idea that there is no friction if the dipoles do not move.

Coming back to the heat propagation equation, due to the boundary conditions 
$\delta \theta(z=0,t) = 0 = \delta \theta(z=e,t)$, Eq.~(\ref{E6}) is solved by decomposition in a series of spatial modes 
labelled by their wave vector $K= m \pi/e$ with $m$ an odd integer. The mode $m=1$ dominates the temperature increase, 
and we keep only the $m=1$ and $m=3$ modes since it is enough to get an accuracy of the order of $1\%$. By averaging spatially 
these two modes, we obtain the $\delta T(t)$ to be used in Eq.~(\ref{E4}):

\begin{eqnarray}
 \delta T_0 & = & \delta T^\star \left(1+\frac{1}{3^4} \right) \hbox{\ with\ } \delta T^\star = \frac{\epsilon_{0} \chi_{1}'' \omega E^2 e^2}{24 \kappa_{th}}  \nonumber \\
 \delta T_{2}(t) & = & \delta T^\star \left( \frac{\cos(2 \omega t-\phi_{2,a})}{\sqrt{1+(2 \omega \tau_{th})^2}}  + \frac{\cos(2 \omega t-\phi_{2,b})}{3^4 \sqrt{1+(2 \omega \tau_{th}/3^2)^2}} \right),
\label{E7}
\end{eqnarray}

where $\tau_{th}= ce^2/(\kappa_{th} \pi^2)$ and $\phi_{2,a} = \phi + \arctan(2 \omega \tau_{th})$ are involved in the dominant $m=1$ mode, while $\phi_{2,b} = \phi + \arctan(2 \omega \tau_{th}/3^2)$ appears in the much less important $m=3$ spatial mode evoked above \cite{precisions}. 

Following Eqs.~(\ref{E4})-(\ref{E5}), the two terms arising in $\delta T_2(t)$ in Eq.~(\ref{E7}) give a contribution to $\chi_{3,h}$. 
These contributions are added, yielding the overestimated value of $\chi_{3,h}$, as well as the damped value of $\chi_{3,h}$ (after multiplying by the function $R(\omega \tau)$ given 
in Eq.~(\ref{E12})). Then, as explained above, Eq.~(\ref{E13}) is used to convert these values in terms of a contribution to  $X_3$.

\section{\label{III} Results}

\subsection{\label{IIIA} Behavior at low and high temperature}

We shall first get some insight into the heating contribution $X_{3,h}$ to the total nonlinear normalized susceptibility $X_3$
by extracting from the previous equations its frequency and thickness dependences. From Eqs.~(\ref{E4}),~(\ref{E5}),
~(\ref{E7}), we keep the leading term to obtain: 

\begin{equation}\label{E15}
\vert X_{3,h}\vert \sim \left\vert \frac{\partial \chi_1}{\partial T} \right\vert 
\frac{\chi_{1}'' \omega e^2}{\sqrt{1+(\omega /\omega_{th})^2}} , 
\end{equation}

where we have defined $\omega_{th}= 2 \pi f_{th} = 1/(2 \tau_{th})$. In Eq.~(\ref{E15}) two characteristic frequencies 
appear: $f_{\alpha}$, which strongly depends on $T$, and the thermal frequency which is inversely proportional to the 
thickness of the sample. For a given experiment, and thus a given $e$, we have to distinguish two regimes: the 
low temperature regime where $f_{\alpha} < f_{th}$ and the high temperature regime where $f_{\alpha} > f_{th}$. Besides, 
for glycerol, which is the liquid of interest here, the frequency dependence of $\chi_{1}''$ is
$\chi_{1}'' \sim f/f_{\alpha}$ below $f_{\alpha}$ and $\chi_{1}'' \sim (f/f_{\alpha})^{-0.55}$ above $f_{\alpha}$. 
Last, $\vert \partial \chi_{1}/ \partial T \vert \sim (f/f_{\alpha})^{0.9}$ below $f_{\alpha}$ and 
$\vert \partial \chi_{1}  / \partial T \vert \sim (f/f_{\alpha})^{-0.6}$ above $f_{\alpha}$. This allows to draw 
from Eq.~(\ref{E15}) the frequency and thickness dependencies of $\vert X_{3,h}\vert$.

\begin{figure*}
\hskip -11mm
\includegraphics[scale=1.0,angle=0]{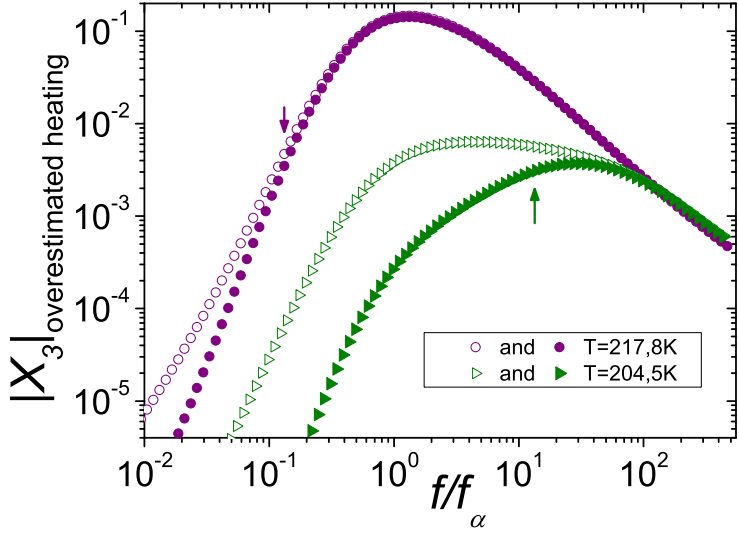}
\caption{\label{Fig1} (Color on line) 
Modulus of the overestimated heating contribution $X_{3,h}$ for $T=204.5$K (triangles) and 
$T=217.8$K (circles), for glycerol and $e=19\mu$m, as a function of the frequency normalized to the 
relaxation frequency $f_{\alpha}(T)$. The open symbols are for a `one sample experiment': the very flat maximum of 
$\vert X_{3,h}\vert$ at $204$K is typical of the `low temperature regime' (see text), while the 
`sharp' maximum close to $f=f_{\alpha}$ at $217.8$K characterizes the `high temperature regime' (see text). 
The filled symbols correspond to the `two samples bridge' described in Refs. \cite{Crauste10,Thibierge08}, 
where the heating contribution of the two samples, $e_{thin}=19\mu$m and $e_{thick}=41\mu$m, cancel each other 
at low enough frequency, i.e. for $f \ll f_{th}(e_{thick}) = 65$Hz, see text. The net heating contribution
is thus reduced in this setup: at $204.5$K this reduction is so strong that the maximum is shifted up slightly above $65$Hz (the up arrow corresponds to $65$Hz at $204.5$K) . 
At $217.8$K the weaker reduction is present only below $65$Hz (see the down arrow).}
\end{figure*}

\begin{figure*}
\hskip -11mm
\includegraphics[scale=1.0,angle=0]{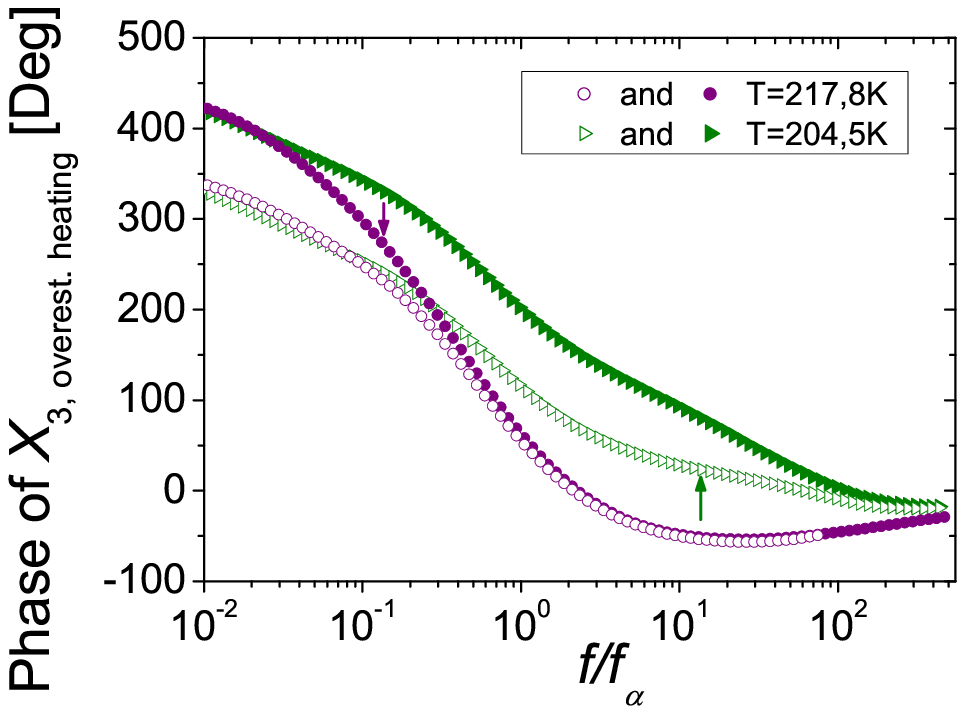}
\caption{\label{Fig2}  (Color on line) Phase of the overestimated heating contribution $X_{3,h}$ for $T=204.5$K 
(triangles) and $T=217.8$K (circles), for glycerol and $e=19\mu$m, as a function of the frequency normalized to the 
relaxation frequency $f_{\alpha}(T)$. The open symbols are for a `one sample experiment'.
The filled symbols correspond to the `two samples bridge' technique of Refs. \cite{Crauste10,Thibierge08}, 
where the heating contribution of the two samples, $e_{thin}=19\mu$m and $e_{thick}=41\mu$m, cancel each other 
at low enough frequency, i.e. for $f \ll f_{th}(e_{thick}) = 65$Hz, see text. The net heating contribution
in this setup thus depends on $f/f_{th}(e_{thick})$. At $204.5$K, $f \simeq f_{th}(e_{thick})$ 
corresponds to $f/f_{\alpha} \simeq 13$, see the up arrow, hence the separation between the filled and the open triangles ends only for 
the highest frequencies. Accordingly, at $217.8$K, the corresponding separation arises at $f/f_{\alpha} \simeq 0.3$, not far from $f \simeq f_{th}(e_{thick})$ which  
corresponds to $f/f_{\alpha} \simeq 0.13$, see the down arrow. 
In the limit $f\ll f_{th}(e_{thick})$, the shift of the phase
in the two samples setup with respect to the one sample setup is $\pi/2$.}
\end{figure*}

\textit{(i)} In the low temperature regime, one gets :

\begin{eqnarray}
\vert X_{3,h}\vert & \sim & f^{2.9} e^2     \hbox{\ when \ } 0 \le f \le f_{\alpha} \nonumber \\
\vert X_{3,h}\vert & \sim & f^{-0.15} e^2 \hbox{\ when \ } f_{\alpha} \le f \le f_{th} \nonumber \\
\vert X_{3,h}\vert & \sim & f^{-1.15} e^0 \hbox{\ when \ } f_{th} \le f . 
\label{E16}
\end{eqnarray}

In Eq.~(\ref{E16}), the maximum over frequency of $\vert X_{3,h}\vert$ arises when the exponent 
of $f$ changes its sign: this maximum value is thus proportionnal to $e^2$, i.e. the heating contribution 
vanishes in the limit of very thin samples. This shows the non intrinsic (or 
spurious) nature of the heating contribution which is studied in this paper. We note that for $e \simeq 19 \mu$m, 
which corresponds to the experiment of Ref.~\cite{Crauste10}, 
one typically gets $f_{th}= 300$Hz. Since the standard definition of 
the glass transition temperature $T_g$ corresponds to $f_{\alpha}(T_g)=0.01$Hz, the 
glass transition temperature is deeply in the low temperature regime. Therefore for $e \simeq 19 \mu$m, 
the heating contribution disappears close enough to $T_g$. This would not be true for samples where the thickness 
$e$ is millimetric.

\textit{(ii)} In the high temperature regime, one gets :

\begin{eqnarray}
\vert X_{3,h}\vert & \sim & f^{2.9} e^2     \hbox{\ when \ } 0 \le f \le f_{th} \nonumber \\
\vert X_{3,h} \vert & \sim & f^{1.9} e^0 \hbox{\ when \ } f_{th} \le f \le f_{\alpha} \nonumber \\
\vert X_{3,h} \vert & \sim & f^{-1.15} e^0 \hbox{\ when \ } f_{\alpha} \le f .
\label{E17}
\end{eqnarray}

Eq.~(\ref{E17}) shows that the maximum over frequency of $\vert X_{3,h}\vert$ arises for 
$f = f_{\alpha}$ and  is proportional to $e^0$, i.e. independent of the thickness of the sample, just as the
 intrinsic non linear response. With $f_{th} \simeq 300$Hz, this high temperature regime onsets at $T \ge 216$K, 
nearly $10$K below the maximum temperature studied in Ref.~\cite{Crauste10}. We thus expect that 
$\vert X_{3,h}\vert$ could play a role for the highest temperatures reported in \cite{Crauste10}. 
The open circles of Fig.~\ref{Fig1} show the `overestimated value' of $\vert X_{3,h}\vert$ for $T=217.8K$. 
The exponents predicted in Eq.~(\ref{E17}), for $f \le f_{th}$ and $f \ge f_{\alpha}$, are well observed. 
Besides, by taking into account the prefactor
not explicitly written in Eq.~(\ref{E17}), one can check that the maximum over frequency of $\vert X_{3,h}\vert$ 
should arise for $f \simeq f_{\alpha}$ and give a value of order $0.15$, as observed on Fig.~\ref{Fig1}.
The comparison to the experimental data will be presented in section~(\ref{IIIC}).

\subsection{\label{IIIB} Heating contribution cancellation at low frequency with the bridge technique}

Before moving to the detailed study of $X_{3,h}(\omega,T)$, we investigate here the consequences of
the fact that our experiment reported in Ref.~\cite{Crauste10} was performed with a bridge 
technique. Two samples of different thicknesses, $e_{thin} \simeq 19 \mu$m and $e_{thick} \simeq 41 \mu$m were used 
in a bridge to suppress the $3\omega$ voltage due to the voltage source imperfection and to the 
(small) non linearity of the voltage detector \cite{Thibierge08,Crauste10}. 
Once the bridge is equilibrated, the voltages $V_{applied, 1 \omega}$ applied 
to each of the two samples are stricly equal. The field applied onto the thin sample is thus larger than 
that applied onto the thick sample by a factor $e_{thick}/e_{thin}$. The subtraction operated by the bridge 
 does not cancel the sought nonlinear response, since the latter goes as $E^3$. 
We show now that this `two samples technique' \textit{strongly reduces} the values of 
$\vert X_{3,h} \vert$ as long as $f \ll f_{th}$. Indeed, from Eq.~(\ref{E7}), one finds, in the limit $f \ll f_{th}$:

\begin{equation}\label{E18}
\vert \delta T_2 \vert = \frac{\epsilon_{0} \chi_{1}'' \omega E^2 e^2}{24 \kappa_{th}} 
\sim \left(V_{applied, 1 \omega}\right)^2 , 
\end{equation}

 i.e. the value of $\delta T_2$ is the same for the thin and the thick sample. By using Eqs.~(\ref{E4}),(\ref{E5}), this 
implies that the two heating contributions in the bridge setup 
perfectly cancel each other in the limit $f \ll f_{th}$. 

This is illustrated in Fig.~\ref{Fig1} where the filled symbols are the result corresponding to 
the bridge technique, while the open symbols are for a `one sample' experiment. At $T=204$K, one sees that
$\vert X_{3,h}\vert$ is strongly reduced by the bridge technique as long as $f \le 13 f_{\alpha}$ which amounts to 
$f \le f_{th}(e_{thick}) \simeq 65$Hz : the small remaining $\vert X_{3,h}\vert$ comes from 
corrections which depend on $f/f_{th}$ to Eq.~(\ref{E18}), and Fig.~\ref{Fig2} reveals that the leading correction, in the limit 
 $f \ll f_{th}$, mainly produces a $\pi /2$ shift of the phase of $X_{3,h}$. On the contrary, for $f \gg f_{th}$, one sees on Figs.~\ref{Fig1}-\ref{Fig2} that the heating contributions are similar, both in phase and magnitude, for a bridge setup and a `one sample' experiment. 
 
The same happens at $T=218$K where $f \le f_{th}(e_{thick}) = 65$Hz
 corresponds again to the frequency range where the bridge technique reduces (and phase shifts) the heating 
contribution. Since the distinction between the `high' and `low' temperature regimes introduced in section~(\ref{IIIA}) 
involves a comparison between $f_{th}$ and $f_{\alpha}$, one concludes that the reduction of the heating 
contribution will be very important in the low temperature regime and much less important in the high temperature regime. 
More precisely, in a one sample experiment, for the low temperature range, the heating contribution has a very flat maximum around 
a few times $f_{\alpha}$. The example of the $204$K curve of Fig.~\ref{Fig1} shows that the two samples setup reduces the heating 
contribution so strongly that its maximum over frequency is shifted slightly above $f_{th}(e_{thick}) = 65$Hz, i.e. at
 $f \gg f_{\alpha}$. For the same reasons, the bridge technique extends slightly upwards the low temperature regime: 
 the high temperature regime, characterised by a well defined maximum of $X_{3,h}$ located 
 around $f=f_{\alpha}$, only arises at $T \ge 218$K instead of the $216$K value evoked in the section~(\ref{IIIA}) where 
 the two samples technique was not taken into account.
 
In the following, all the results shown in Figs.~\ref{Fig3}-\ref{Fig10} correspond to a `two samples' setup, in order 
to be directly comparable to our results of Ref. \cite{Crauste10}.

\subsection{\label{IIIC} Main features of the heating contribution to $\chi_3$}

\begin{figure*}
\hskip -11mm
\includegraphics[scale=1.0,angle=0]{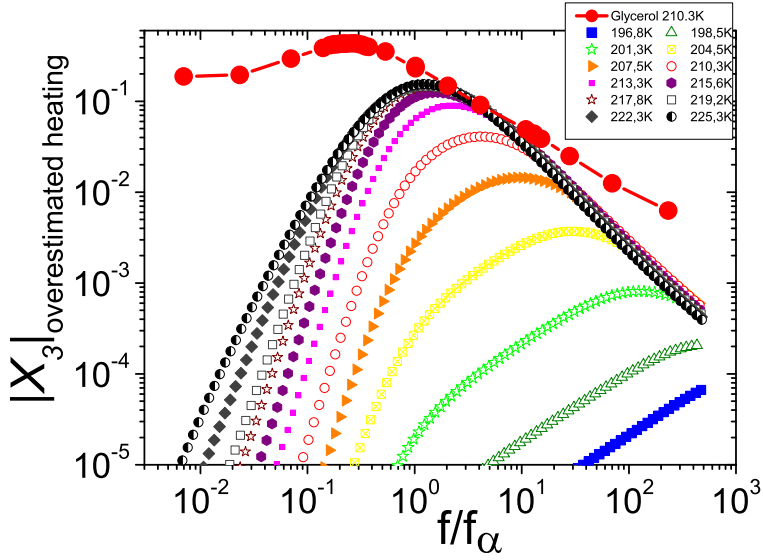}
\caption{\label{Fig3}  (Color on line) Modulus of the overestimated heating contribution $\vert X_{3,h} \vert$ for the two 
glycerol samples setup $e_{thin}=19\mu$m, $e_{thick}=41\mu$m of Ref. \cite{Crauste10}, as a function of the frequency normalized to the 
relaxation frequency $f_{\alpha}(T)$. For comparison 
the measured values of $\vert X_{3}\vert$ \cite{Crauste10} are given for $T=210.3$K (filled red circles). 
Several features of this overstimated 
heating contribution are at odds with those measured in \cite{Crauste10}, e.g. the maximum over frequency 
increases with $T$, it occurs at a temperature dependent value of $f/f_{\alpha}$ and TTS is not obeyed (see text). }
\end{figure*}

\begin{figure*}
\hskip -11mm
\includegraphics[scale=1.0,angle=0]{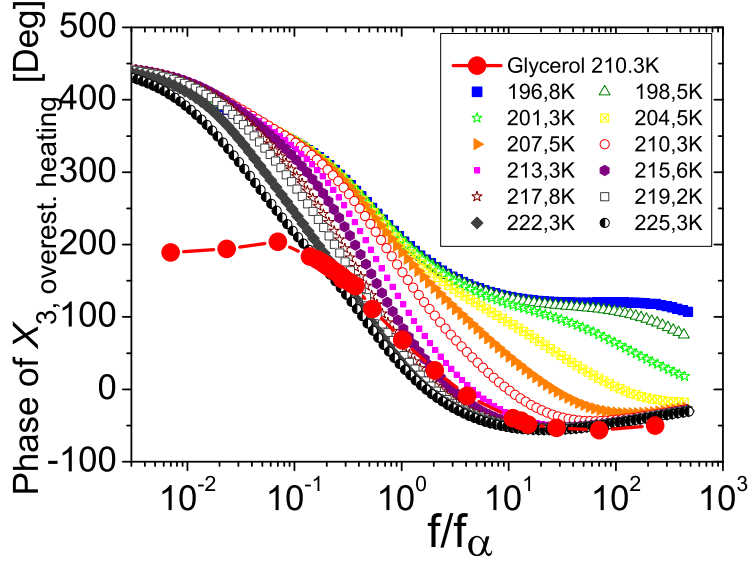}
\caption{\label{Fig4}  (Color on line) Phase of the overestimated heating contribution $X_{3,h}$ for the two 
glycerol samples setup $e_{thin}=19\mu$m, $e_{thick}=41\mu$m of reference \cite{Crauste10}, as a function of the frequency 
normalized to the relaxation frequency $f_{\alpha}(T)$. For comparison 
the measured values of $X_3$ at $T=210.3$K \cite{Crauste10} are given (filled red circles): they do not significantly vary with $T$ 
since the data reported in \cite{Crauste10} obey TTS (see text), contrarily to the heating contribution phases.}
\end{figure*}

\begin{figure*}
\hskip -11mm
\includegraphics[scale=1.0,angle=0]{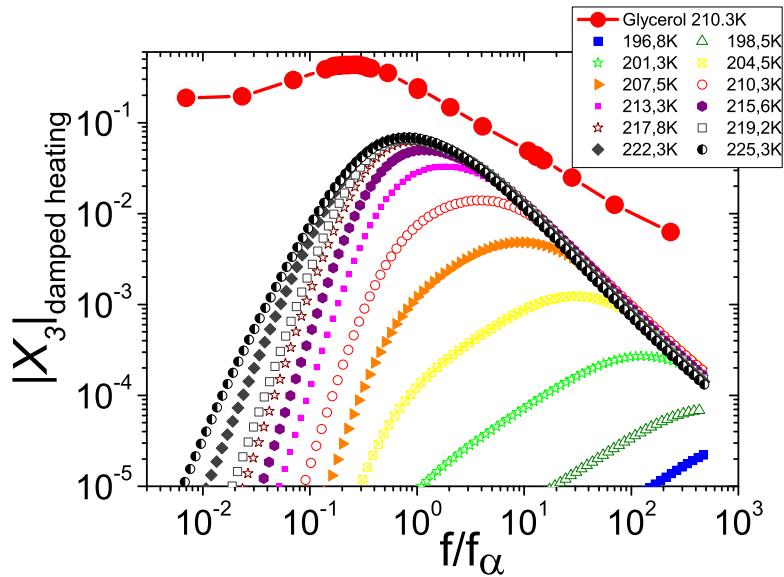}
\caption{\label{Fig5}  (Color on line) Modulus of the damped heating contribution $X_{3,h}$ for the two 
glycerol samples setup $e_{thin}=19\mu$m, $e_{thick}=41\mu$m of Ref. \cite{Crauste10}, as a function of the frequency 
normalized to the relaxation frequency $f_{\alpha}(T)$. For comparison 
the measured values of $X_3$ at $T=210.3$K \cite{Crauste10} are given (filled red circles). The comments made in the caption of
 Fig.~\ref{Fig3} still apply in this case, with the noticeable difference that the values of 
$\vert X_{3,h}\vert$ are now smaller than those of Fig.~\ref{Fig3} due to the fact that one takes into account 
the damping effect arising from the finite relaxation time of the dipoles (see text section~\ref{IIA2} and Eq.~(\ref{E12})) .}
\end{figure*}

\begin{figure*}
\hskip -11mm
\includegraphics[scale=1.0,angle=0]{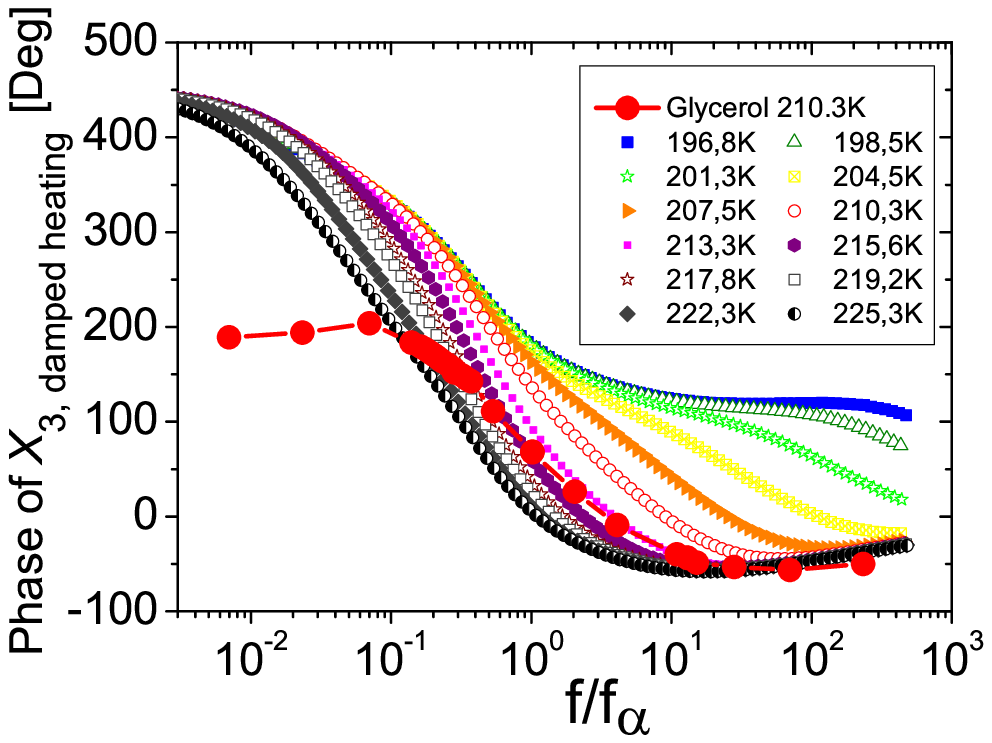}
\caption{\label{Fig6}  (Color on line) Phase of the overestimated heating contribution $X_{3,h}$ for the two 
glycerol samples setup $e_{thin}=19\mu$m, $e_{thick}=41\mu$m of reference \cite{Crauste10}, as a function of the frequency 
normalized to the relaxation frequency $f_{\alpha}(T)$. The comment of the caption 
of Fig.~\ref{Fig4} applies also here.}
\end{figure*}

We shall now discuss the main features of the overestimated heating contribution (Figs.~\ref{Fig3},\ref{Fig4}), 
as well as those of the damped heating contribution (Figs.~\ref{Fig5}-\ref{Fig6}). They will be compared to
the experimental $X_{3}$ ($\simeq N_{corr}{\cal H}$)  values at $T=210.3$K presented in Ref. \cite{Crauste10}. 
Two properties reported in \cite{Crauste10} are of interest here. First, when the temperature $T$ is varied between 
$225.3$K and $194.0$K, the maximum value $max_f(\vert X_{3} \vert)$ over frequency of the measured 
$\vert X_{3} \vert$ increases by a factor $\simeq 1.5$. Second, in this temperature interval, the 
frequency dependence of $\vert X_{3} \vert$ at each temperature, once rescaled vertically by its $T$-dependent 
$max_f(\vert X_{3} \vert)$ value, fall onto a master curve depending only of $f/f_{\alpha}$. Besides, the phase of $ X_3$ also depends only of $f/f_{\alpha}$. This `Time Temperature Superposition' (TTS) property of $X_{3}$ is analogous to the similar properties of other well studied observables, e.g. $\chi_1$, for many supercooled liquids. Considering the fact that the above mentioned $1.5$ factor is modest 
with respect to the nearly five orders of magnitudes displayed in Figs.~\ref{Fig3},\ref{Fig5}, we are led to the conclusion that the 
$T$ variations of $\vert X_{3} \vert$  are hardly visible in these two plots.

From Figs.~\ref{Fig3}-\ref{Fig6}, one gets the five following features, common both to the overstimated and damped heating contributions:

\textit{(i)} For a given $f/f_{\alpha}$, $X_{3,h}$ mainly increases with $T$, at odds with the
 experimental behavior reported in Ref. \cite{Crauste10}. Besides the magnitude of the $T$-dependence of $X_{3,h}$ is much larger than that of $X_{3}$ reported in Ref. \cite{Crauste10}.

\textit{(ii)} The results for the modulus and the phase show that the heating contributions do not obey TTS. For the reasons 
explained in section~\ref{IIIB}, the maximum over frequency of $X_{3,h}$ arises for $f \gg f_{\alpha}$ at low temperature, and for 
$f \simeq f_{\alpha}$ in the $218$K$-225$K interval. We remind here that the maximum value over frequency
of the experimental $\vert X_{3} \vert$ $\simeq$ $\vert N_{corr} {\cal H} \vert$ arises at a very different frequency $f \simeq 0.21 f_{\alpha}$ 
 -see Ref. \cite{Crauste10}-. 
Summarising points \textit{(i)-(ii)}, the 
 frequency and $T$ dependence of the heating contributions do not, at all, look like those of $X_{3}$.

\textit{(iii)} For $f \simeq 0.21 f_{\alpha}$, the heating contribution is always negligible with respect to the values of 
$X_{3}$ reported in Ref. \cite{Crauste10}.  More precisely, the heating correction to $max_f(\vert X_{3} \vert)$ 
is smaller than $1.5 \%$ for $T \le 217.8K$, and reaches $5 \%$ for the highest temperature of $225.3$K. In this specific case 
of the highest temperature, subtracting the $5 \%$ heating contribution would \textit{decrease} 
$max_f(\vert X_{3} \vert)$  by an 
amount equal to the error bar given in the Figure 3 of Ref.\cite{Crauste10}. 

\textit{(iv)} For $f \gg f_{\alpha}$, the heating contributions decrease significantly faster with frequency (as $ f^{-1.15}$) than 
$\vert X_{3} \vert$. Thus the influence of the heating contributions disappears in the limit $f \gg f_{\alpha}$.

\textit{(v)} For $T \le 202-204$K, the heating correction is always negligible for any frequency, i.e. a worst case analysis 
shows that it would amount to modify $\vert X_{3} \vert$ by less than the $5 \%$ error bar reported in 
Ref. \cite{Crauste10}. This means that in the range $[194$K$,204$K$]$, the data of Ref. \cite{Crauste10} are absolutely
free of any heating contribution, whatever the frequency.

\subsection{\label{IIID} Evidence of a damping of the heating contribution to $\chi_3$}

In the range $[194$K$,204$K$]$, the absence of heating contribution guarantees that $N_{corr} {\cal H}$, 
obeying TTS, is well measured. In the $[204$K$,225$K$]$ range, the $(N_{corr} {\cal H})_{glyc}$ values of glycerol can be
investigated by subtracting the heating contribution 
to the measured $X_3(f,T)$ reported at various temperatures in \cite{Crauste10}. As we have seen, the heating contribution is weak
or negligible in many cases, but it is interesting to study to what extent the expected TTS is better verified after such a subtraction.
We define :

\begin{equation}\label{E19}
(N_{corr} {\cal H})_{glyc} = X_3 - X_{3,h} . 
\end{equation}

\begin{figure*}
\hskip -11mm
\includegraphics[scale=1.0,angle=0]{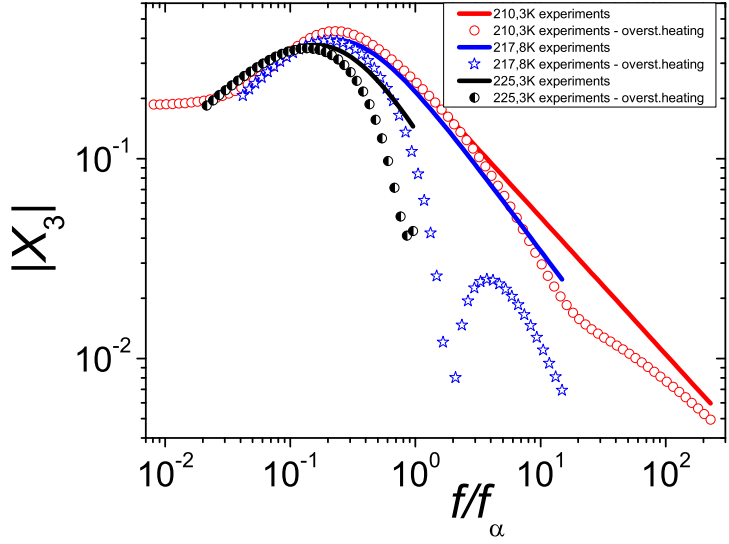}
\caption{\label{Fig7}  (Color on line) Solid lines: interpolated values of the experimental 
$\vert X_3 \vert$ values reported 
in \cite{Crauste10} for glycerol, obtained with the `two samples bridge setup'. The intrinsic $\vert N_{corr} {\cal H} \vert_{glyc}$ of 
glycerol is plot as symbols and results from the complex subtraction $X_3- X_{3,h}$ where the overestimated heating 
contribution of $X_{3,h}$ (displayed in Figs.~\ref{Fig3}-~\ref{Fig4}) is used. The resulting $\vert N_{corr} {\cal H} \vert_{glyc}$ 
curves, which display a dip or a strong variation of the $f$ dependence for the three temperatures considered here, are non 
TTS, contrary to what is expected in the case of no heating effect.}
\end{figure*}

\begin{figure*}
\hskip -11mm
\includegraphics[scale=1.0,angle=0]{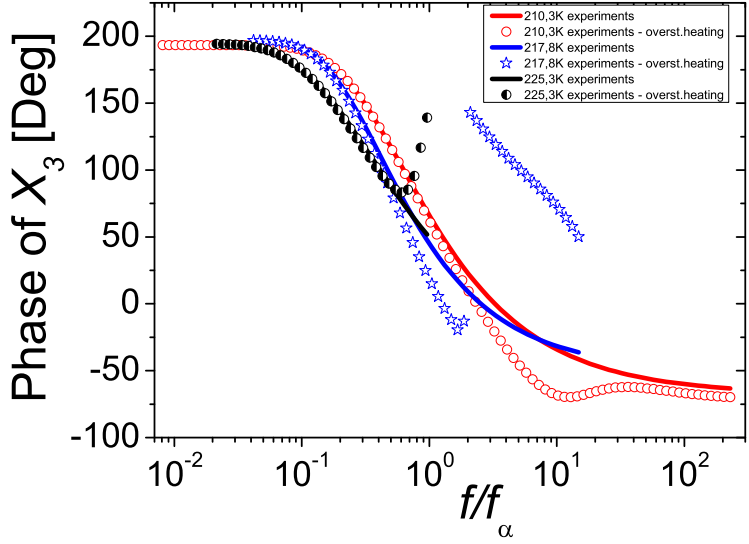}
\caption{\label{Fig8}  (Color on line) 
Solid lines: interpolated values of the phase of the experimental 
$X_3$ reported in \cite{Crauste10} for glycerol, obtained with the `two samples bridge setup'. Symbols: phases corresponding 
to the complex subtraction evoked in the caption of Fig.~\ref{Fig7} (with the same symbols used). 
The $\pi$ jump occuring for $217.8$K around $f/f_{\alpha}=2$ reveals that $\vert N_{corr} {\cal H} \vert_{glyc}$ vanishes
 at this point. }
\end{figure*}

\begin{figure*}
\hskip -11mm
\includegraphics[scale=1.0,angle=0]{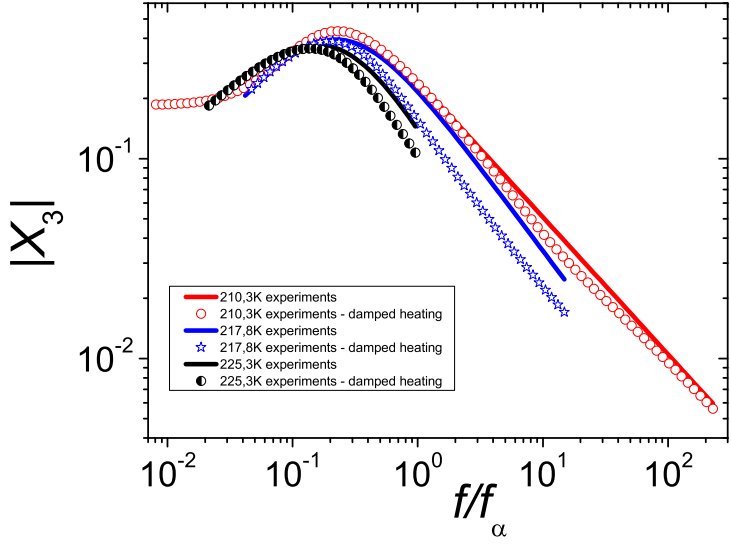}
\caption{\label{Fig9}  (Color on line) (Color on line) Solid lines: interpolated values of the experimental 
$\vert X_3 \vert$ values reported 
in \cite{Crauste10} for glycerol, obtained with the `two samples bridge setup'. The intrinsic $\vert N_{corr} {\cal H} \vert_{glyc}$ of 
glycerol is plot as symbols and results from the complex subtraction $X_3- X_{3,h}$ where the damped heating 
contribution of $X_{3,h}$ (displayed in Figs.~\ref{Fig5}-~\ref{Fig6}) is used. The dips visible in Fig.~\ref{Fig7} 
are no longer visible, and the resulting $\vert N_{corr} {\cal H} \vert_{glyc}$ are approximately TTS. 
This suggests that the damped heating estimate is reasonable.}
\end{figure*}

\begin{figure*}
\hskip -11mm
\includegraphics[scale=1.0,angle=0]{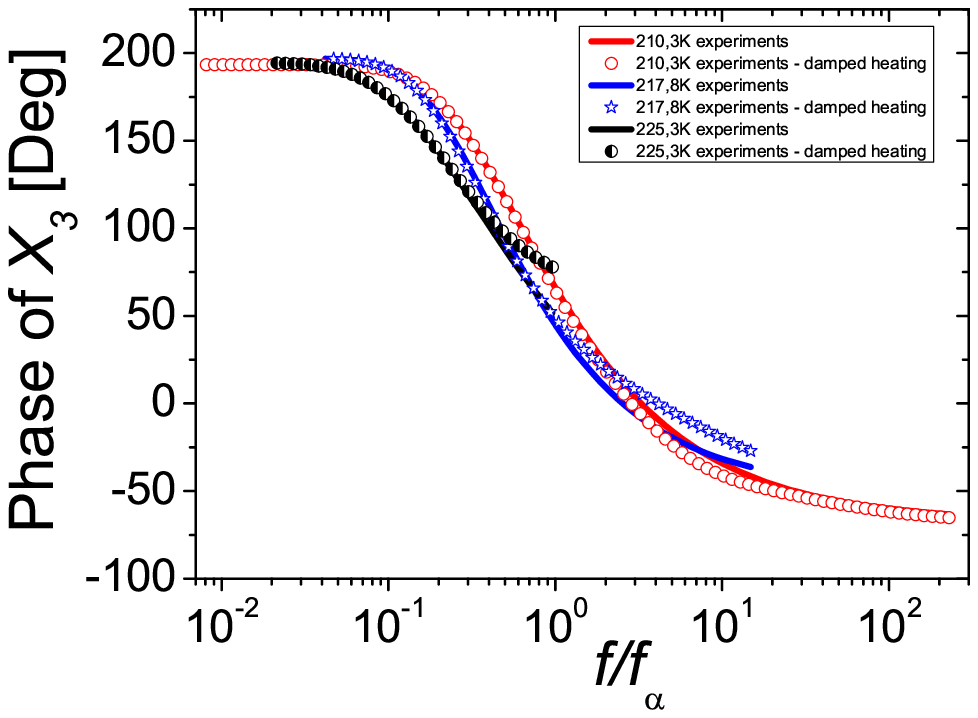}
\caption{\label{Fig10}  (Color on line) 
Solid lines: interpolated values of the phase of the experimental 
$X_3$ reported in \cite{Crauste10} for glycerol, obtained with the `two samples bridge setup'. Symbols: phases corresponding 
to the complex subtraction evoked in the caption of Fig.~\ref{Fig9} (with the same symbols used). 
The phase of the resulting $(N_{corr} {\cal H})_{glyc}$ values are not significantly 'less TTS' than the original ones. 
This suggests that the damped heating estimate, displayed in Figs.~\ref{Fig5}-~\ref{Fig6}, is reasonnable.}
\end{figure*}

We performed this (complex) subtraction for the three temperatures above $204$K in Ref. \cite{Crauste10}, 
i.e. $210.3$K, $217.8$K, $225.3$K. For each of these three temperatures, the frequency dependence is first 
fitted by a smooth function interpolating between the frequencies: 
these three fits appear as continuous lines in Figs.~\ref{Fig7}-\ref{Fig10}. Then, the subtraction of 
$X_{3,h}$ to the measured $X_3$ is performed, $X_{3,h}$ being 
either the overestimated contribution(see Figs.~\ref{Fig7},~\ref{Fig8}) or the damped one (see Figs.~\ref{Fig9},~\ref{Fig10}). 
The results of this subtraction appear as open symbols in Figs.~\ref{Fig7}-\ref{Fig10}. 

When the overestimated value of $X_{3,h}$ is used, $(N_{corr} {\cal H})_{glyc}$ may differ strongly from the measured 
$X_3$. Fig.~\ref{Fig7} shows that the shape of the curve giving the modulus of 
$(N_{corr} {\cal H})_{glyc}$ vs frequency is modified at $f/f_{\alpha} \simeq 20$ 
for $T$ = $210.3$K. At $217.8$K, a dip appears around $f/f_{\alpha} \simeq 2$, and at $225.3$K the dip is present at 
$f/f_{\alpha} \simeq 0.8$. These features come from the frequency range where the values of $\vert X_{3,h}\vert$ 
are close to or larger than those of $\vert X_{3} \vert$, as shown in Fig.~\ref{Fig3}. Depending on the 
difference between the phases of  $X_{3}$ and of $X_{3,h}$, the effect on 
$\vert N_{corr} {\cal H} \vert _{glyc}$ is more or less pronounced: at $T=217.8K$ the dip at 
$f/f_{\alpha} \simeq 2$ corresponds to a \textit{vanishing} value of $\vert N_{corr} {\cal H} \vert _{glyc}$, as shown by the $\pi$ jump 
on the phase (see Fig.~\ref{Fig8}). At $210.3$K and $225.3$K, the non zero value of 
$\vert N_{corr} {\cal H} \vert _{glyc}$ is related to a less pronounced effect on the phase (see Fig.~\ref{Fig8}). 
To summarize, using 
the overstimated values of $X_{3,h}$ yields non TTS features in the resulting $(N_{corr} {\cal H})_{glyc}$. 
We note that this is true already for the $210.3$K curve corresponding to a temperature close to $204.7$K
for which the curve of Ref. \cite{Crauste10} verifies TTS.
The fact that our experimental $X_3(f,T)$ curves verify TTS is a strong indication of the overestimated character 
(already anticipated) of the heating contribution used to calculate the results presented in Figs.~\ref{Fig7}-~\ref{Fig8}. 
We are thus led to the conclusion that the heating contribution has to be damped. 

We now perform the same analysis as above, using the damped $X_{3,h}(f,T)$ values displayed in Figs.~\ref{Fig5}-~\ref{Fig6}. 
At first glance, Figs.~\ref{Fig9}-\ref{Fig10} reveal that for the damped heating estimate,
the complex subtraction no longer produces the strong non TTS
features depicted above. The resulting $\vert N_{corr} {\cal H} \vert _{glyc}$ does not strongly differ from the original 
$ \vert X_3\vert$: the value of the exponent giving the decay at $f/f_{\alpha} > 0.3$ of 
$\vert N_{corr} {\cal H} \vert _{glyc}$ is only $5\%$ larger than the corresponding one for $\vert X_3 \vert$ at $217.8K$ 
 and $225.3$K. The above mentionned feature at $210.3$K and $f/f_{\alpha} \simeq 20$ has almost 
disappeared. Last, the phase of $(N_{corr} {\cal H}) _{glyc}$ is not significantly `less TTS' 
than the original phase of $X_3$. We thus conclude that the damped $X_{3,h}$ values, introduced 
in section (\ref{IIA2}) and displayed in Figs~\ref{Fig5}-~\ref{Fig6}, meet the requirement resulting from the last 
point of section (\ref{IIIC}), namely that $(N_{corr} {\cal H})_{glyc}$ should be TTS. This is an experimental indication that the 
damping factor introduced in Eq.~(\ref{E12}) is reasonnable, despite the two simplifying assumptions made along its derivation.

We leave for future work the very difficult experimental task of isolating the heating contribution in itself. Two main ideas 
could be conceivable. First, one could try to push the $225K$ experiment in the range$f/f_{\alpha} \gg 1$ where some non TTS 
feature might arise on the phase of $(N_{corr} {\cal H})_{glyc}$ if one extrapolates the calculations of Fig.~\ref{Fig10}. In practice it 
is  extremely difficult to measure accurately $X_{3}$ in this range of parameters, because of the d.c. heating contribution 
coming from the whole thermal circuit between each sample and the experimental cell \cite{precision2}. This d.c. heating 
contribution does not cancel in our two samples' bridge, and thus the balance condition changes at each value 
of the voltage source. The second idea would be to use the fact that the canceling effect of the heating contribution in the bridge arises 
only for $f \le f_{th}(e_{thick})$. As a result, the heating contribution to $X_3$ should be different for two setups, 
say $A$ and $B$, where $e_{thick}$ is different. In the frequency 
range $f_{th}(e_{thick, A}) \le f \le f_{th}(e_{thick, B})$, one may hope that the differences between the values of 
$X_{3,h}$ could be detectable. Putting numbers, in the case $e_{thick,A}=41\mu$m and 
$e_{thick,B}=25\mu$m (see \cite{Epaps10}), shows that this difference does not reach $10\%$ of $X_3$ (even for the optimal frequency), and is therefore very difficult to single out unambiguously.

\section{\label{IV} Conclusion}

We have presented a thorough study of the `homogeneous' 
heating contribution to the third harmonics experiments carried out on 
glycerol in Ref. \cite{Crauste10} between $T_g+4$K and $T_g+35$K. We have emphasized the `spurious' nature of this 
heating contribution by showing that it vanishes for thin enough samples and low enough temperatures. We have shown that 
the `two samples' bridge technique, presented in Ref. \cite{Crauste10}, widens the temperature interval over which the heating 
contribution is totally negligible with respect to the measured $N_{corr} {\cal H}$ values reported in \cite{Crauste10}. Globally, 
the heating contribution exhibits behaviors very different from those of the nonlinear normalized susceptibility $X_{3}$ 
reported in \cite{Crauste10}: It (mainly) increases with 
$T$ for a given $f/f_{\alpha}$, its peak arises at much higher frequencies than in \cite{Crauste10}, its frequency dependence 
is faster, it does not obey TTS (except for the highest temperatures). At a quantitative level, we have shown that one can 
safely neglect the  heating contribution to the maximum over frequency of $\vert X_3 \vert$ for all the 
temperature interval studied in \cite{Crauste10}. Thanks to the quite large temperature interval where the absence of any 
heating contribution is guaranteed, 
we have shown that the intrinsic $N_{corr} {\cal H}$ of glycerol obeys TTS in this temperature range. 
Extrapolating this TTS feature up to $225K$ allows to put an experimental constraint on the homogeneous 
heating contribution. We obtain the following important result: The fact that
the homogeneous heating contribution must be damped because of the finite relaxation time of the dipoles is confirmed
by an investigation of the shape and TTS property of the $X_3(f,T)$ curves. In addition, we may conclude that the damping factor 
given by Eq.~(\ref{E12}) is reasonnable.


\begin{thebibliography}{99}
\bibitem{Langevin} The Debye dynamics can be also obtained by starting with the non inertial Langevin equation for the rotational Brownian motion of a particule, by appropriate transformation of the variables and direct averaging of the stochastic equation so obtained.
\bibitem{justification} Since the relative thermal variation of $\tau$ is much larger than that of $\Delta \chi_1$, the latter quantity is taken as a constant when solving Eq. \ref{E8}.
\bibitem{precision2} The approximation that the electrodes are a perfect thermostat are in general valid for computing 
$\delta T_2$, see \cite{electrodes}, but it is in general not valid for computing the d.c. temperature increase. 
The difference lies in the fact that, because of the `skin effect', the temperature oscillations are efficiently damped in the 
thick electrodes, while the d.c. component is not damped at all. The d.c. heat flux is thus fully transmitted by the electrodes, 
and the unavoidable insulating piece between the electrode and the experimental cell is responsible for the major part of the 
d.c. temperature increase. The only way to get rid of the d.c. heating is to perform extremely fast experiments, in the spirit 
of what is made in the seminal reference \cite{Richert06}, so as to measure the sample properties in a time scale much lower than 
the thermal diffusion time between the sample and the experimental cell.
\end{thebibliography}
\end{document}